\documentclass[aps,twocolumn,prb,amsmath,amssymb]{revtex4}

\usepackage{graphicx,epsfig}
\usepackage{color}
\usepackage{wasysym}

\begin{document}
\title{Heat-charge mixed noise and thermoelectric efficiency fluctuations}
\author{Adeline Cr\'epieux$^{1}$}
\author{Fabienne Michelini$^{2}$}
\affiliation{$^1$Aix Marseille Universit\'e, Universit\'e de Toulon, CNRS, CPT UMR 7332, 13288 Marseille, France}
\affiliation{$^2$Aix Marseille Universit\'e, CNRS, IM2NP, UMR 7334, 13288 Marseille, France}

\begin{abstract}
The close relationship between the noise and the thermoelectric conversion is studied  in a quantum dot using a quantum approach based on the non-equilibrium Green function technique. We show that both the figure of merit and the efficiency can be written in term of noise and we highlight the central role played by the correlator between the charge current and the heat current that we call the {\it mixed noise}. After giving the expression of this quantity as an integral over energy, we calculate it, first in the linear response regime, next in the limit of small transmission through the barriers (Schottky regime) and finally in the intermediate regime. We discuss the notion of efficiency fluctuations and we also see here that the mixed noise comes into play.
\end{abstract}

\maketitle


\section{Introduction}

The adage {\it ``the noise is the signal''} enunciated by Landauer \cite{Landauer98} has been proved many
times. In the domain of electric transport for example, zero-frequency noise gives the dc conductance in the linear
regime and also gives access to the charge of the carriers in the weak transmission regime via the Schottky relation, whereas finite-frequency noise gives the ac conductance via the fluctuation-dissipation theorem. One of
the beautiful illustrations of this was the measurement
of the fractional charge of a two dimensional electron gas in the
fractional quantum Hall regime \cite{Saminadayar97,Picciotto97}. There are numerous studies on charge current noise in quantum system, both theoretically \cite{Lesovik89,Buttiker90,Beenakker91,Buttiker92,Blanter00,Martin05} and experimentally \cite{Li90,Dekker91,Liefrink91,Reznikov95,Brom99,Deblock03,Zakka07,Basset10,Basset12,Parmentier12,Altimiras14}, which remains a very active field. Moreover, in the recent wave of quantum heat transport studies, several
works are devoted to the statistic of the heat \cite{Kindermann04,Saito11,Kumar12,Sanchez12,Golubev13}, to the heat current noise both at zero-frequency \cite{Saito07,Battista13} and finite-frequency \cite{Averin10,Sergi11,Zhan11}, and to the heat fluctuations in driven coherent conductors \cite{Battista14a,Moskalets14}.

So far very few works have been devoted to the
correlations between the charge and heat currents \cite{Giazotto06,Sanchez13} that we call
the {\it mixed noise}. This lacuna needs to be filled, even through there are some recent works in that direction \cite{Crepieux15a,Crepieux15b,Battista14b}. In addition to the characterization of such a ``new'' quantity, our objective was to find which kind of
information can be extracted from the mixed noise. For this, we have calculated
the correlator between the charge current and the heat current for a
two terminal quantum dot system using the non-equilibrium Green function technique and
we have studied its possible relation with quantities such as the
thermoelectric efficiency and the figure of merit. Indeed, it turns out that this quantity plays an important role in the quantification of the thermoelectric conversion which is a hot topic 
nowadays \cite{Whitney14,Whitney15,Whitney15b,Entin15,Dare15}.

This paper is organized as follows:  in Sec.~II, we present the system, we give the definition of the mixed noise and we describe the method to calculate it for a quantum dot. We give its expression in Sec.~III, we discuss the conservation rules and we give the explicit expressions of the figure of merit and efficiency, first in the linear response regime (Sec.~III.A), second in the Schottky regime (Sec.~III.B) and third in the intermediate regime (Sec.~III.C). We consider the issue of the thermoelectric efficiency definition for a fluctuating system in Sec.~IV, and conclude in Sec.~V.


\section{System, definition and method}

\begin{figure}[!h]
\begin{center}
\includegraphics[width=8cm]{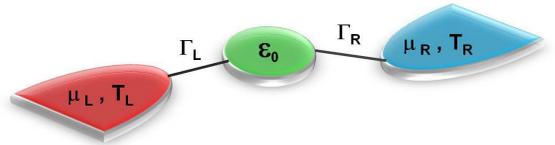}
\caption{Schematic picture of a single level quantum dot
connected to two reservoirs with distinct chemical potentials $\mu_{L,R}$ and
temperatures $T_{L,R}$. The voltage and temperature gradients are respectively given by $eV=\mu_L-\mu_R$, and $\Delta T=T_L-T_R$. We also define the average temperature $T_0=(T_L+T_R)/2$.}
\label{figure1}
\end{center} 
\end{figure}

We have a single level quantum dot connected to two
reservoirs with distinct chemical potentials $\mu_{L,R}$ and temperatures
$T_{L,R}$ as depicted on Fig.~\ref{figure1}. The Hamiltonian contains four parts: $H=H_L+H_R+H_\mathrm{dot}+H_T$, where $H_{p=L,R}=\sum_{k\in p}\varepsilon_kc^\dag_kc_k$ is the energy of the left~(L) and right~(R) reservoirs, $H_\mathrm{dot}=\varepsilon_0d^\dag d$ is the energy of the dot, and $H_T=\sum_{p=L,R}\sum_{k\in p}V_kd^\dag c_k+h.c.$ corresponds to the transfer of one electron from the reservoirs to the dot and vice versa. $c^\dag_k$ ($d^\dag$) and $c_k$ ($d$) are respectively the creation and annihilation operators in the reservoirs (dot).

In a similar way to the standard definitions of the charge current
noise, $\mathcal{S}_{pq}^{II}$, and heat current noise, $\mathcal{S}_{pq}^{JJ}$, we define the mixed
noise as the zero-frequency Fourier transform of the correlator
mixing the charge current in the $p$ reservoir and the heat current in the $q$ reservoir:
\begin{eqnarray}\label{def_mixed_noise}
\mathcal{S}_{pq}^{IJ}=\int_{-\infty}^\infty\langle \delta\hat I_p(0)\delta\hat J_q(t)\rangle dt~,
\end{eqnarray}
where $\delta\hat I_p(t)=\hat I_p(t)-\langle \hat I_p\rangle$ and $\delta\hat J_p=\hat J_p(t)-\langle \hat J_p\rangle$ are the charge and heat current deviations from their averages. The charge and heat current operators \cite{Crepieux12} are defined through
$\hat I_p(t)=-e \dot N_p(t)$ and \cite{note1},
\begin{eqnarray}
\label{heat_current_def}
\hat J_p(t)&=&\hat I^E_p(t)-(\mu_p/e) \hat I_p(t)~,
\end{eqnarray}
with $\hat N_p$, the electrons number operator and $\hat I^E_p=-dH_p/dt$, the energy current operator, both associated to the reservoir $p$. A direct calculation gives:
\begin{eqnarray}
\hat I_p(t)&=&\frac{ie}{\hbar}\sum_{k\in p}(V_kc^\dag_k d-V_k^*d^\dag c_k )~,
\end{eqnarray}
and,
\begin{eqnarray}
\label{heat_current}
\hat J_p(t)&=&\frac{i}{\hbar}\sum_{k\in p}(\varepsilon_k-\mu_p)(V_kc^\dag_k d-V_k^*d^\dag c_k )~.
\end{eqnarray}
When one inserts these expressions of charge and heat current operators in Eq.~(\ref{def_mixed_noise}), one obtains a correlator with four operators, i.e., a two-particles Green function, mixing the creation and annihilation operators of reservoirs and quantum dot, which can be expressed in terms of the two-particles Green function of the dot only. Next, using decoupling\cite{haug07} which applies for a non-interacting system (via the Wick's theorem), one can rewrite the mixed noise in terms of single-particle dot Green functions.

\section{Mixed noise}

Assuming to be in the wide-band limit and performing a Fourier transform in order to replace the time integration by an energy integration, we finally obtain the expression of the zero-frequency mixed
noise \cite{Crepieux15a} in terms of transmission coefficient $\mathcal{T}$ and Fermi-Dirac
distribution function $f_p$:
\begin{eqnarray}\label{mixed_noise}
\mathcal{S}_{pq}^{IJ}=(2\delta_{pq}-1)\frac{e}{h}\int_{-\infty}^\infty (\varepsilon-\mu_q)\mathcal{F}(\varepsilon)d\varepsilon~,
\end{eqnarray}
with
\begin{eqnarray}\label{fct_F}
\mathcal{F}(\varepsilon)&=&\mathcal{T}(\varepsilon)\Big[f_L(\varepsilon)\big[1-f_L(\varepsilon)\big]+f_R(\varepsilon)\big[1-f_R(\varepsilon)\big]\Big]\nonumber\\
&&+\mathcal{T}(\varepsilon)\big[1-\mathcal{T}(\varepsilon)\big]\big[f_L(\varepsilon)-f_R(\varepsilon)\big]^2~,
\end{eqnarray}
and $f_{p}(\varepsilon)=[1+\exp((\varepsilon-\mu_p)/(k_BT_p))]^{-1}$.
The expression of the mixed noise given by Eq.~(\ref{mixed_noise}) resembles the expression of the charge noise: $\mathcal{S}_{pq}^{II}=(2\delta_{pq}-1)e^2h^{-1}\int_{-\infty}^\infty \mathcal{F}(\varepsilon)d\varepsilon$, with an additional factor $(\varepsilon-\mu_q)$. We recall that for the heat noise, we have: $\mathcal{S}_{pq}^{JJ}=(2\delta_{pq}-1)h^{-1}\int_{-\infty}^\infty (\varepsilon-\mu_p) (\varepsilon-\mu_q)\mathcal{F}(\varepsilon)d\varepsilon$.

In the absence of any time-dependent excitation, i.e., in the stationary regime, the total number of electrons is conserved: $\langle \dot N_L\rangle+\langle \dot N_R\rangle=0$ which leads immediately to the cancellation of the total charge current: $\langle \hat I_L\rangle+\langle \hat I_R\rangle=0$. Such a cancellation does not apply for the total heat current, indeed from Eq. (\ref{heat_current}) we have:
\begin{eqnarray}\label{total_heat_current}
\langle \hat J_L\rangle+\langle \hat J_R\rangle&=&-\langle\dot E_L\rangle-\langle\dot E_R\rangle+\mu_L\langle\dot N_L\rangle+\mu_R\langle\dot N_R\rangle\nonumber\\
&=&(\mu_R-\mu_L)\langle \hat I_L\rangle/e=V\langle \hat I_R\rangle~,
\end{eqnarray}
since $\langle\dot E_L\rangle+\langle\dot E_R\rangle=0$ (energy conservation), where $V$ is the bias voltage between the left and the right reservoirs. The total heat current is non-zero in the junction because of heat dissipation (Joule effect) at the contact resistances between the dot and the reservoirs. However, Eq.~(\ref{total_heat_current}) is consistent with the power conservation: $\langle \hat P^\mathrm{th}\rangle=\langle \hat P^\mathrm{el}\rangle$, since it is an equality between the average values of the thermal power, $\hat P^\mathrm{th}=\hat J_L+\hat J_R$, and the electric power, $\hat P^\mathrm{el}=V\hat I_R$.

We turn now our interest to the fluctuations conservation. Taking the double sum on the reservoirs, we obtain
$\sum_{p,q\in[L,R]}\mathcal{S}_{pq}^{IJ}=\sum_{p,q\in[L,R]}\mathcal{S}_{pq}^{II}=0$
and,
\begin{eqnarray}
\label{heat_noise}
&&\sum_{p,q\in[L,R]}\mathcal{S}_{pq}^{JJ}=V^2\mathcal{S}_{pq}^{II}~.
\end{eqnarray}
The total mixed noise cancels as it is the case for the total charge noise, whereas the total heat noise does not cancel. However, Eq.~(\ref{heat_noise}) can be interpreted as a power fluctuations conservation. Indeed it can be rewritten as an equality between the thermal power fluctuations and the electric power fluctuations, that is:
\begin{eqnarray}
\int_{-\infty}^\infty\langle \hat P^\mathrm{th}(t)\hat P^\mathrm{th}(0)\rangle dt=\int_{-\infty}^\infty\langle \hat P^\mathrm{el}(t)\hat P^\mathrm{el}(0)\rangle dt~.
\end{eqnarray}

In the following, we exploit the result given by Eq.~(\ref{mixed_noise}) by studying its limits in three distinct regimes: first, in the linear regime for voltage and temperature gradients much smaller than the other characteristic energies of the sample (linear response), second, far from equilibrium for a weak transmission coefficient (Schottky regime), and third, in the intermediate regime.

\subsection{Linear response}

Bulk thermoelectric devices operate mostly in the  linear response regime. In this regime, the charge current is the same in both reservoirs in absolute value (we remove the reservoir index) and is proportional to the voltage and temperature gradients. The same applies for the heat current. Thus, the charge and heat currents are given by the matrix equation:
\begin{eqnarray}
\left(
\begin{array}{c}
\langle \hat I\rangle\\
\langle \hat J\rangle
\end{array}
\right)=
\left(
\begin{array}{ccc}
G&&GS\\
\Pi G&&K+\Pi S G
\end{array}
\right)
\left(
\begin{array}{c}
V\\
\Delta T
\end{array}
\right)~,
\end{eqnarray}
where $G$ and $K$ are the electric and thermal conductances, $S$ and $\Pi$ are the Seebeck and Peltier coefficients. These two last quantities obey the Onsager relation, $\Pi=ST_0$, where $T_0$ is the average temperature of the sample. To derive the maximum of efficiency of the system, one needs first to express the voltage and the heat current in term of the charge current and temperature gradient:
\begin{eqnarray}
\left(
\begin{array}{c}
V\\
\langle \hat J\rangle
\end{array}
\right)=
\left(
\begin{array}{ccc}
G^{-1}&&-S\\
S T_0&&K
\end{array}
\right)
\left(
\begin{array}{c}
\langle \hat I\rangle\\
\Delta T
\end{array}
\right)~.
\end{eqnarray}
The macroscopic efficiency \cite{Esposito15} of a thermoelectric device is defined as the ratio between the average output power and the average input power:
\begin{eqnarray}\label{def_eta}
\eta_M=\frac{\langle \hat P^\mathrm{out}\rangle}{\langle \hat P^\mathrm{in}\rangle}~,
\end{eqnarray}
where $\langle \hat P^\mathrm{in}\rangle$ and $\langle \hat P^\mathrm{out}\rangle$ are given either by the thermal power or by the electrical power depending  on the thermoelectric device one considers, i.e., electric generator or refrigerator. To find the maximum of efficiency, we look for the zero of the derivative of Eq.~(\ref{def_eta}) according to the charge current. It leads to the equation
$ST_0 \langle \hat I\rangle^2+2K \Delta T \langle \hat I\rangle-SK G \Delta T^2=0$,
whose solution reads as:
\begin{eqnarray}
\langle \hat I\rangle=\frac{K\Delta T}{ST_0}\left(\sqrt{1+ZT_0}-1\right)~,
\end{eqnarray}
where the figure of merit is defined by $ZT_0=S^2T_0G/K$. When inserted in the expression of the efficiency given by Eq.~(\ref{def_eta}), this allows us to write the maximum of efficiency under the form:
\begin{eqnarray}
\eta_\mathrm{max}=\eta_C\frac{\sqrt{1+ZT_0}- 1}{\sqrt{1+ZT_0}+ 1}~,
\end{eqnarray}
with $\eta_C=\Delta T/T_0$, the Carnot efficiency.

In the linear response regime, we have $eV\approx k_B\Delta T\approx 0$ which gives $f_L(\varepsilon)\approx f_R(\varepsilon)$: the second line in Eq.~(\ref{fct_F}) is thus negligible, whereas the terms of the first line in Eq.~(\ref{fct_F}) lead to contributions for the noises that are proportional to the derivative of the charge or of the heat currents according either to the voltage or to the temperature gradient. This leads to the following relation between the 
noises and the conductances:
\begin{eqnarray}
\mathcal{S}^{II}&=&2k_BT_0G~,\nonumber\\\label{FDT}
\mathcal{S}^{IJ}&=&\mathcal{S}^{JI}=-2k_BT_0^2SG~,\\
\mathcal{S}^{JJ}&=&2k_BT_0^2(K+S^2T_0G)~.\nonumber
\end{eqnarray}
These results are in agreement with the fact that the fluctuation-dissipation
theorem holds for every type of noise. Note that the $p$ and $q$
indexes have been removed in Eq.~(\ref{FDT}) since the absolute values of the noises are identical
in amplitude in both reservoirs in the linear response regime.
From these results, one can extract the expressions of the conductances in term of noises:
\begin{eqnarray}
G&=&\frac{\mathcal{S}^{II}}{2k_BT_0}~,\nonumber\\
S&=&-\frac{\mathcal{S}^{IJ}}{T_0\mathcal{S}^{II}}~,\\
K&=&\frac{1}{2k_BT_0^2}\left(\frac{\mathcal{S}^{II}\mathcal{S}^{JJ}-(\mathcal{S}^{IJ})^2}{\mathcal{S}^{II}}\right)~.\nonumber
\end{eqnarray}
Finally, the thermoelectric figure of
merit can be expressed fully in terms
of noises such as \cite{Crepieux15a}:
\begin{eqnarray}\label{ZT0}
ZT_0=\frac{(\mathcal{S}^{IJ})^2}{\mathcal{S}^{II}S^{JJ}-(\mathcal{S}^{IJ})^2}~.
\end{eqnarray}
Note that thanks to the Cauchy-Swartz inequality, we have $(\mathcal{S}^{IJ})^2\le \mathcal{S}^{II}S^{JJ}$, which ensures a positive and non-diverging $ZT_0$ (provided that $K$ is non-zero). Moreover, a vanishing mixed noise $\mathcal{S}^{IJ}$ leads to the cancellation of the figure of merit and of the efficiency.

\subsection{Schottky regime}

By reducing the dimension of the thermoelectric device, there is a great chance to leave the linear response regime, in particular when one operates in the limit of a weak transmission through the barriers, $\mathcal{T}(\varepsilon)\ll 1$. In that case, the noises are proportional to the currents (Schottky regime). Indeed, one can neglect the $\mathcal{T}^2$ contribution in Eq.~(\ref{fct_F}) and show that:
\begin{eqnarray}\label{V0}
\mathcal{S}^{II}_{LR}&=&\mathcal{C}e\langle \hat I_R\rangle~,\\\label{V1}
\mathcal{S}^{IJ}_{LR}&=&\mathcal{C}e\langle \hat J_R\rangle=\mathcal{C}(\varepsilon_0-\mu_R)\langle \hat I_R\rangle~,\\\label{V2}
\mathcal{S}^{JJ}_{LR}&=&\mathcal{C}(\varepsilon_0-\mu_L)\langle \hat J_R\rangle~,
\end{eqnarray}
where $\mathcal{C}=\coth[(\varepsilon_0-\mu_R)/2k_BT_R-(\varepsilon_0-\mu_L)/2k_BT_L]$ is a temperature factor which is equal to the unit at $T_{L,R}=0$, and:
\begin{eqnarray}
\langle \hat I_{L,R}\rangle &=&\pm\frac{e}{h}\int_{-\infty}^\infty [f_L(\varepsilon)-f_R(\varepsilon)]\mathcal{T}(\varepsilon)d\varepsilon~,\\
\langle \hat J_{L,R}\rangle &=&\pm\frac{1}{h}\int_{-\infty}^\infty (\varepsilon-\mu_R)[f_L(\varepsilon)-f_R(\varepsilon)]\mathcal{T}(\varepsilon)d\varepsilon~.\nonumber\\
\end{eqnarray}
Thus, in the same way and for the same reason that the charge noise is proportional to the charge current, the mixed noise is proportional to the heat current.

Far from the linear response regime, the figure of merit
is no longer the relevant parameter to quantify thermoelectric
conversion and one has rather to come back to the efficiency definition, i.e., the ratio between the averages of  output and input powers.
According to the thermoelectric engine that one wants to build,
the powers are given either by the product between charge current
and voltage or directly by the heat current. In general, it is not
possible to connect these quantities to the noises. However, in the
Schottky regime, i.e., in the weak transmission regime, the currents can be written as (see Eqs.~(\ref{V0}) and (\ref{V1})):
\begin{eqnarray}\label{I}
\langle \hat I_R\rangle &=&(\mathcal{C}e)^{-1}\mathcal{S}^{II}_{LR}~,\\\label{J}
\langle \hat J_R\rangle&=&(\mathcal{C}e)^{-1}\mathcal{S}^{IJ}_{LR}~.
\end{eqnarray}
By calculating the difference between Eqs.~(\ref{V1}) and (\ref{V2}), one can find the expression of the voltage in term of currents and noises, that is $
V=\mathcal{S}^{IJ}_{LR}/(\mathcal{C}e\langle \hat I_R\rangle)-\mathcal{S}^{JJ}_{LR}/(\mathcal{C}e\langle \hat J_R\rangle)
$. Using Eqs. (\ref{I}) and (\ref{J}), it becomes
$V=\mathcal{S}^{IJ}_{LR}/\mathcal{S}^{II}_{LR}-\mathcal{S}^{JJ}_{LR}/\mathcal{S}^{IJ}_{LR}$.
Finally, the macroscopic efficiency of a thermoelectric refrigerator, $\eta_M=|\langle \hat J_R\rangle/(V\langle \hat I_L\rangle)|$, reads as \cite{Crepieux15a}:
\begin{eqnarray}\label{eta}
\eta_M=\frac{(\mathcal{S}^{IJ}_{LR})^2}{|\mathcal{S}^{II}_{LR}\mathcal{S}^{JJ}_{LR}-(\mathcal{S}^{IJ}_{LR})^2|}~.
\end{eqnarray}

\begin{widetext}

\begin{figure}[h!]
\begin{center}
\includegraphics[width=5.5cm]{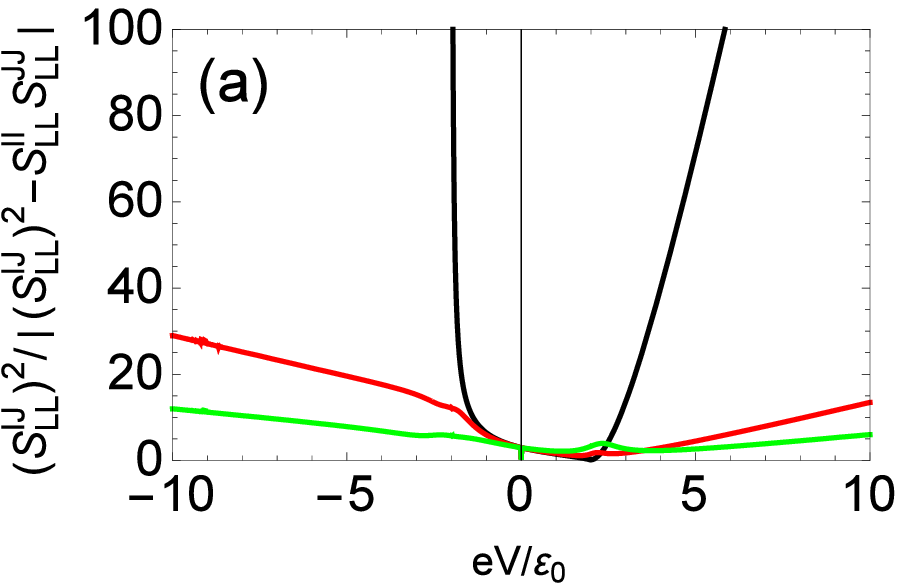}
\includegraphics[width=5.5cm]{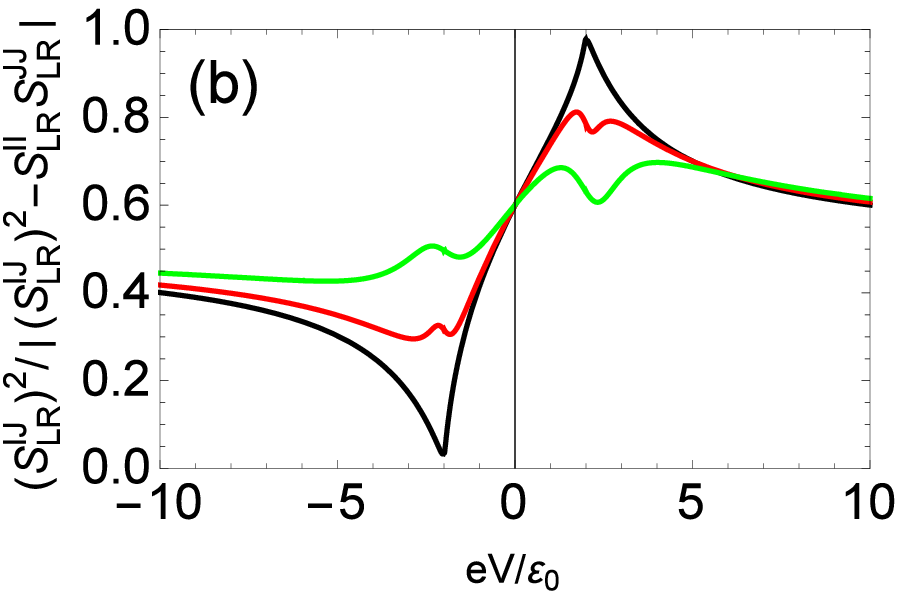}
\includegraphics[width=5.5cm]{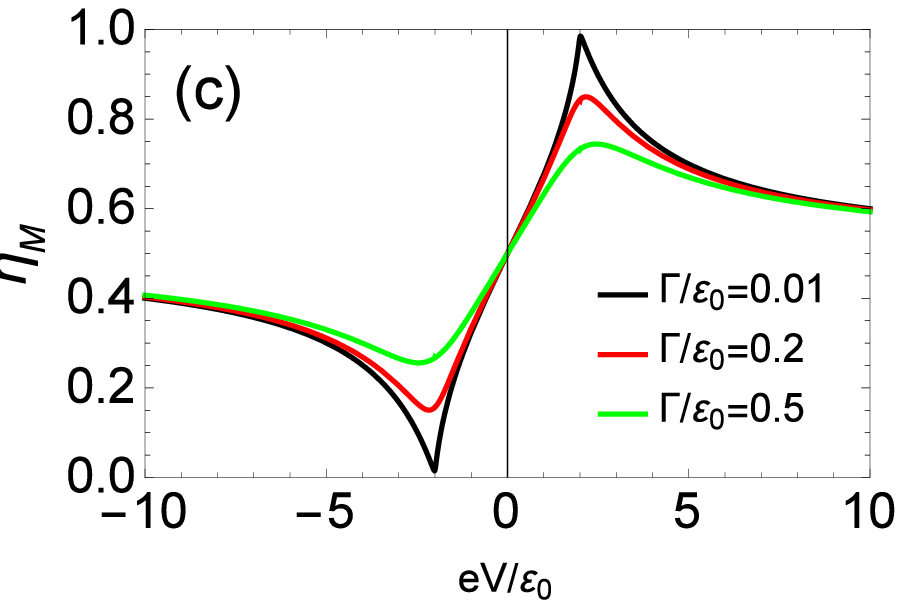}
\caption{(a) Auto-ratio of noises, (b) cross-ratio of noises, and (c) thermoelectric efficiency as a function of voltage $V$ for $\mu_R=-eV/2$, $\mu_L=eV/2$, $k_BT_0/\varepsilon_0=0.001$, and $\Delta T=0$. At weak $\Gamma$ (black curve), the cross-noise ratio and the efficiency fit together, in agreement with Eq.~(\ref{eta}). When $\Gamma$ increases (red and green curves), the profiles of these two quantities differ since the device no longer works in the Schottky regime. The profile of the auto-ratio noise has nothing to do with the two other quantities, even for weak $\Gamma$.}
\label{figure2}
\end{center} 
\end{figure}

\end{widetext}

Note that the temperature factor $\mathcal{C}$ does not appear in the final expression of the efficiency since there is an exact compensation. It means that Eq.~(\ref{eta}) is valid whatever the temperature is, provided that the device still operates in the Schottky regime. A crucial aspect in Eq.~(\ref{eta}) is the fact that the ratio of cross-noises, i.e., the correlators between distinct
reservoirs ($L$ and $R$) must be considered, otherwise the value that we get does not correspond to
the efficiency. As a graphical proof, taking a Breit-Wigner transmission coefficient: $\mathcal{T}(\varepsilon)=\Gamma^2/[(\varepsilon-\varepsilon_0)^2+\Gamma^2]$ with $\Gamma$ the dot energy widening due to the coupling to the reservoirs, three quantities are plotted in
Fig.~\ref{figure2} from left to right: the auto-ratio of noises, the cross-ratio of noises and the
efficiency. As stated by Eq.~(\ref{eta}), the cross-ratio of noises and the
efficiency coincide at low transmission $\mathcal{T}$, i.e., in the Schottky
regime (compare the black curves corresponding to $\Gamma/\varepsilon_0=0.01$ in Fig.~\ref{figure2}(b) and Fig.~\ref{figure2}(c)), whereas the profile of the auto-ratio of noises has nothing
to do with the two other quantities. From this result, one can
conclude that the mixed cross-noise is a measure of the efficiency. Indeed, for a device to convert electricity to heat or vice-versa, the charge current in one reservoir has to show some correlation with the heat current in the other reservoir. From Eq.~(\ref{eta}) we also see that a vanishing mixed cross-noise directly
cancels the thermoelectric efficiency. In Fig.~\ref{figure2}(c), the efficiency is maximal and equal to one for $eV=2\varepsilon_0$. However, it does not guarantee a high output power. Indeed, in the Schottky limit, the charge current and the heat current (and thus the powers) are weak since they are proportional to $\Gamma^2$, which is small  in that regime in comparison to the other characteristic energies. This is the reason why searching of the maximum of efficiency according to the current or voltage is no longer sufficient and one has also to look for the maximal efficiency at given power output \cite{Whitney14,Whitney15} in order to optimize the nano-device.

\subsection{Intermediate regime}

\begin{figure}[h!]
\begin{center}
\includegraphics[width=9cm]{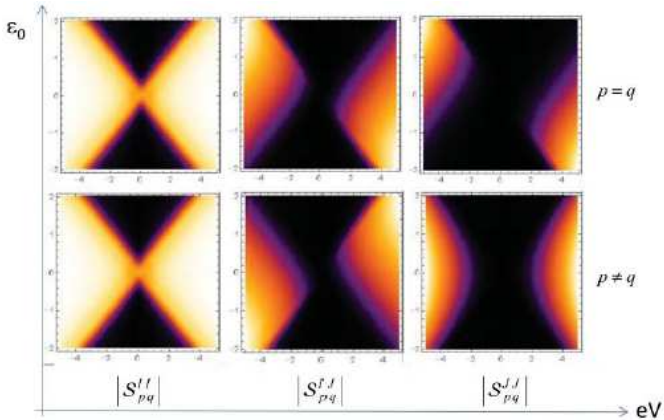}
\caption{Absolute value of the charge noise (left column), mixed
noise (central column) and heat noise (right column) as a function
of $eV/\Gamma$ and $\varepsilon_0/\Gamma$ for $k_BT_{L,R}/\Gamma=1$. The black regions correspond to
the lowest values (close to zero) and the bright ones to the
highest values of the noises.}
\label{figure3}
\end{center} 
\end{figure}

Outside the linear response regime and outside the Schottky regime, i.e.,
in the intermediate regime, one needs to perform numerical
calculations in order to characterize the behavior of mixed
noise since all the energies that enter into play are of the same order of magnitude.
In Fig.~\ref{figure3} are shown the absolute values of the noises as a
function of voltage and dot energy level, taken in the
same reservoir ($p=q$) or in distinct reservoirs ($p\ne q$). We observe that the charge noise keeps the same symmetry whether taken in the same reservoir or in distinct reservoirs (due to the fact that we have $\mathcal{S}^{II}_{p\ne q}=-\mathcal{S}^{II}_{pp}$). This is no longer the case for mixed and heat noises: the mixed noise shows a certain asymmmetry as well as the heat noise taken in the same reservoir. However, the heat noise in distinct reservoirs presents the same symmetry as the charge noise. Thus, when plotting the ratio of noise $(\mathcal{S}^{IJ}_{LR})^2/|\mathcal{S}^{II}_{LR}\mathcal{S}^{JJ}_{LR}-(\mathcal{S}^{IJ}_{LR})^2|$, the asymmetry in voltage and dot energy level results from the mixed noise only.

\begin{figure}[h!]
\begin{center}
\includegraphics[width=4cm]{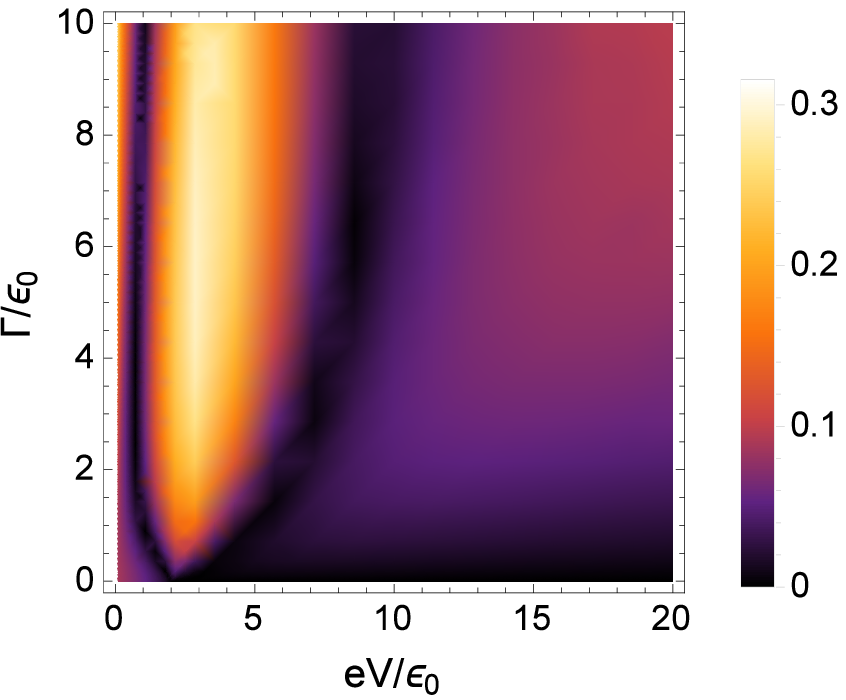}
\includegraphics[width=4cm]{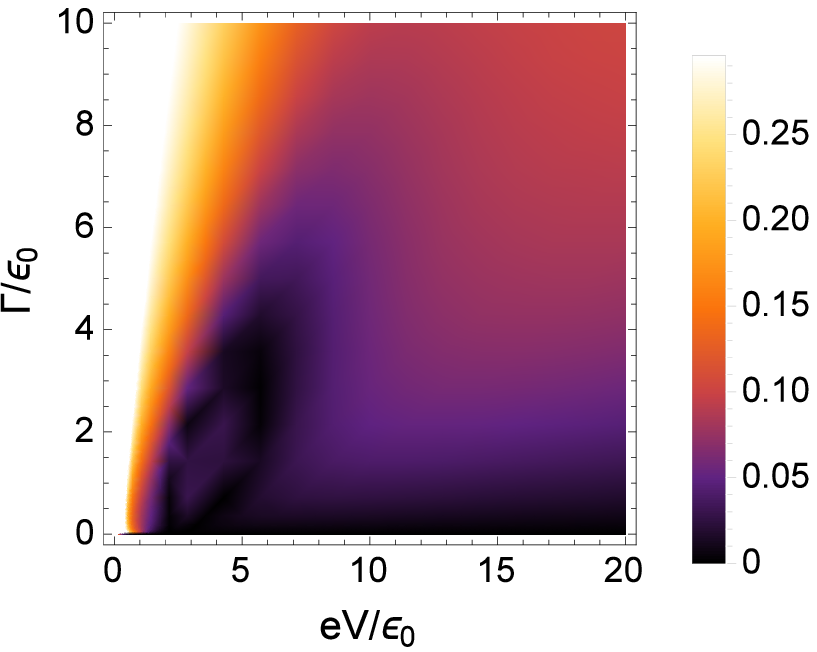}
\caption{Difference in percent between the ratio of noises and the macroscopic efficiency at $k_BT_{L,R}/\varepsilon_0=0.001$ (left graph) and  $k_BT_{L,R}/\varepsilon_0=0.1$ (right graph) as a function of voltage and coupling strength between the dot and the reservoirs.}
\label{figure3bis}
\end{center} 
\end{figure}

Finally, we plot in Fig.~\ref{figure3bis} the difference in percent between the ratio $(\mathcal{S}^{IJ}_{LR})^2/|\mathcal{S}^{II}_{LR}\mathcal{S}^{JJ}_{LR}-(\mathcal{S}^{IJ}_{LR})^2|$ and the efficiency $\eta_M=|\langle \hat J_R\rangle/(V\langle \hat I_L\rangle)|$ as a function of voltage and coupling strength. At low temperature (left graph), three extended black lines on which the ratio and the efficiency coincide  appear, whereas there are one extended line and one loop at higher temperature (right graph). In both case, we observe that the coincidence between the ratio of noises and the efficiency is  not limited to vanishing values of $\Gamma$: it remains valid in regions (appearing in black) that are quite far from the Schottky limit. This strengthens the important role played by the ratio of noises given above.

\section{Efficiency fluctuations}

So far, we have been interested in the relation between the mixed noise and the macroscopic thermoelectric efficiency, neglecting the possible fluctuations of the latter quantity. However, like charge and heat currents, the efficiency is subject to fluctuations and its instantaneous value could overcome during a short time the Carnot efficiency \cite{Esposito15,Verley14a,Verley14b,Campisi14}. In this section, we introduce the notion of efficiency fluctuations and we show that they are related to the mixed noise.

In the presence of fluctuations, the definition of average efficiency is no longer the ratio of the average powers of Eq.~(\ref{def_eta}). In place, we propose to define the efficiency as the time and quantum averages of the output power operator multiplied by the inverse of the input power operator, i.e.
\begin{eqnarray}\label{def_eta_flu}
\eta=\frac{1}{\tau^2}\int_{\;0}^{\;\tau} dt'\int_{\;t'}^{\;\tau} \langle\hat P^\mathrm{out}(t)[\hat P^\mathrm{in}(t')]^{-1}\rangle dt~,
\end{eqnarray}
where $\tau$ is the time measurement and $\hat P^\mathrm{in(out)}(t)=\langle \hat P^\mathrm{in(out)}\rangle+\delta\hat P^\mathrm{in(out)}(t)$. This definition ensures the coincidence of $\eta$ with the macroscopic efficiency when one recovers the classical limit; it makes sense within the assumption of small input power fluctuations avoiding possible divergence. In the steady state, the time translation invariance allows one to simplify Eq.~(\ref{def_eta_flu}) to
\begin{eqnarray}
\eta=\frac{1}{\tau}\int_{\;0}^{\;\tau}\langle\hat P^\mathrm{out}(t)[\hat P^\mathrm{in}(0)]^{-1}\rangle  dt~.
\end{eqnarray}
Let us consider first an electric generator for which $\hat P^\mathrm{in}(0)=\hat J_R(0)$ and $\hat P^\mathrm{out}(t)=V\hat I_L(t)$, assuming that the voltage does not fluctuate. Using $\hat I_L(t)=\langle \hat I_L\rangle+\delta \hat I_L(t)$ and $\hat J_R(0)=\langle \hat J_R\rangle+\delta \hat J_R(0)$, and expanding $[\hat J_R(0)]^{-1}$ up to the first order in $\delta \hat J_R(0)$, we get:
\begin{eqnarray}\label{elec_gene}
\eta=\eta_\mathrm{M}\left(1-\frac{1}{\tau}\int_{\;0}^{\;\tau} \frac{\langle \delta\hat I_L(t)\delta\hat J_R(0)\rangle}{\langle \hat I_L\rangle \langle \hat J_R\rangle}dt\right)~,
\end{eqnarray}
where $\eta_\mathrm{M}=V\langle \hat I_L\rangle/\langle \hat J_R\rangle$ is the macroscopic efficiency \cite{Esposito15}. If one considers instead a refrigerator, the input and output powers are respectively $\hat P^\mathrm{in}(0)=V\hat I_L(0)$ and $\hat P^\mathrm{out}(t)=\hat J_R(t)$. Expending $[\hat I_L(0)]^{-1}$ up to first order in $\delta \hat I_L(0)$, we get:
\begin{eqnarray}\label{refrig}
\eta=\eta_\mathrm{M}\left(1-\frac{1}{\tau}\int_{\;0}^{\;\tau} \frac{\langle \delta\hat J_R(t)\delta\hat I_L(0)\rangle}{\langle \hat I_L\rangle \langle \hat J_R\rangle}dt\right)~,
\end{eqnarray}
where $\eta_\mathrm{M}=\langle \hat J_R\rangle/(V\langle \hat I_L\rangle)$.

\begin{figure}[h!]
\begin{center}
\hspace*{-0.3cm}\includegraphics[width=6.4cm]{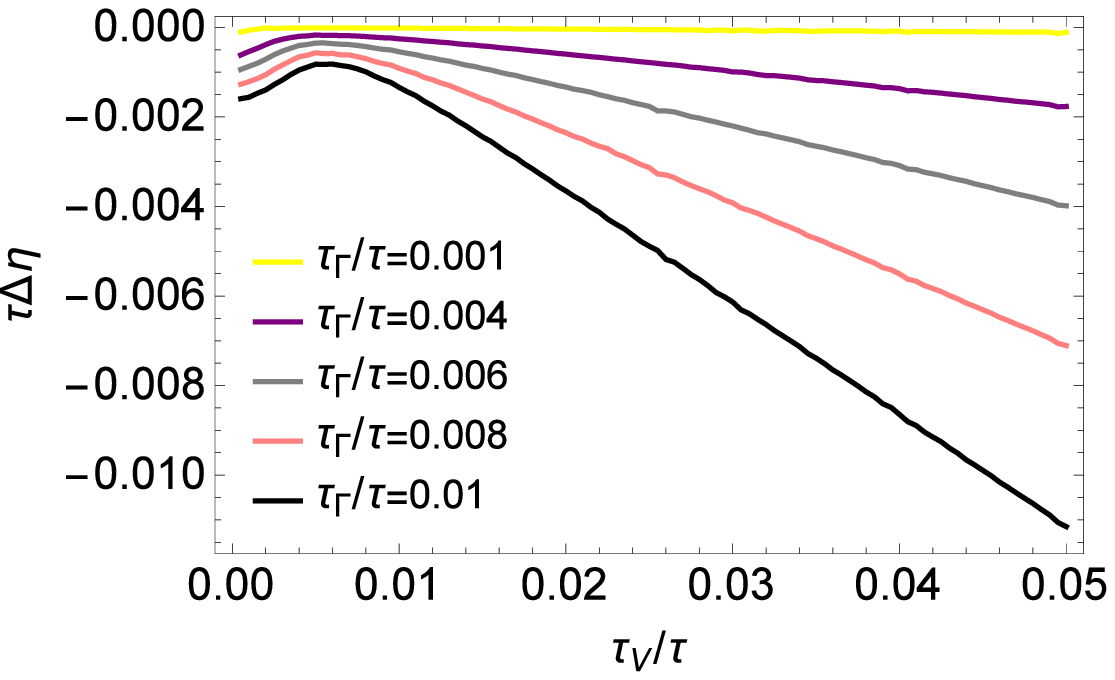}
\includegraphics[width=6cm]{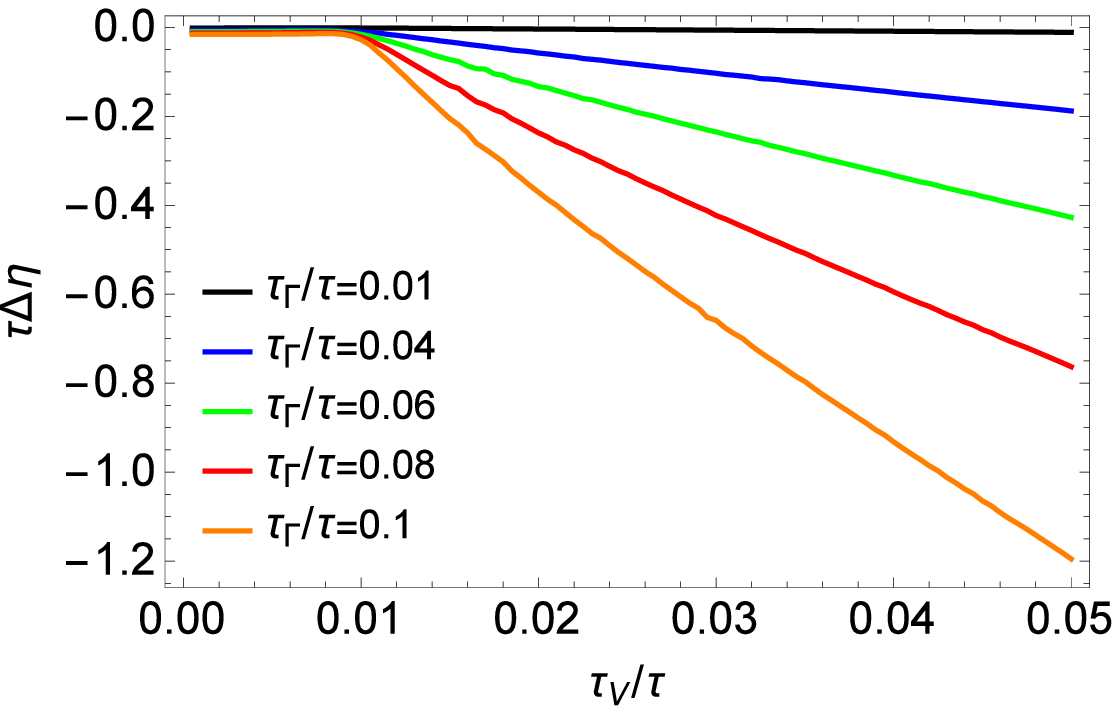}
\caption{The product $\tau\Delta \eta$ as a function of the ratio $\tau_V/\tau$ at $\tau_{\varepsilon_0}/\tau=0.02$ and $T_{L,R}=0$ for varying values of the tunneling time $\tau_\Gamma/\tau$ : in the interval $[0.001,0.01]$ for the top graph and in the interval $[0.01,0.1]$ for the bottom graph.}
\label{figure4}
\end{center} 
\end{figure}

For both thermoelectric engines, the correction to the macroscopic efficiency can be written as an integral over time of the mixed correlator. According to the sign of the second term in Eqs.~(\ref{elec_gene}) and~(\ref{refrig}), the efficiency could be either enhanced or reduced by the fluctuations compared to the macroscopic efficiency. We study numerically the behavior of the efficiency change $\Delta\eta=\eta-\eta_M$ in the limit of a time measurement $\tau$ being the largest characteristic time. Indeed, in that limit, the efficiency change resembles the mixed noise of Eq.~(\ref{def_mixed_noise}) divided by the product of the charge and heat currents. For this purpose, we introduce the tunneling time $\tau_\Gamma=\hbar/\Gamma$  between the dot and the reservoirs as well as the time $\tau_V=\hbar/eV$. We plot on Fig.~\ref{figure4} the product $\tau\Delta\eta$ as a function of $\tau_V/\tau$ for varying values of $\tau_\Gamma/\tau$, keeping all these ratio much smaller than one in order to obey the starting assumption. It is remarkable to observe that $\tau\Delta\eta$
shrinks with increasing coupling strength (decreasing tunneling time), in agreement with the fact that the efficiency has to coincide with the macroscopic efficiency when one reaches the classical regime. In contrast, for decreasing coupling strength (increasing tunneling time) which correspond to a weak transmitting quantum dot, $\tau\Delta\eta$ increases and the efficiency moves away from its macroscopic value. 
Thus, the use of macroscopic efficiency to quantify the thermoelectric conversions is questionable in the quantum regime. Indeed, in this latter case, due to fluctuations, a corrective term is present that can be rewritten for long measurement as the ratio between the mixed noise and the product of heat and charge currents.


\section{Conclusion}

Mixed noise has fulfilled much of its promises. Indeed, we
have shown that: (i) this quantity  is related to the thermoelectric figure of merit in the
linear response regime, (ii) it is related to the thermoelectric efficiency
in the Schottky regime, and (iii) it is related to the efficiency fluctuations.
Thus, this quantity deserves to be studied on the same level as the charge and heat noises, both
theoretically and experimentally. From an experimental perspective, the
challenge is to find a way to measure such a quantity, whereas from
a theoretical point of view, mixed noise needs to be calculated using more realistic
approaches that include, among others, many channels and interactions with several baths.

\acknowledgments

A.C. thanks P. Eym\'eoud, T. Martin and R. Whitney for discussions on noise
and thermoelectricity.



\begin{thebibliography}{20}
\bibitem{note1}Taking the average value of Eq.~(\ref{heat_current_def}), we get $\dot Q_p=\dot E_p-\mu_p \dot N_p$, 
where $Q_p$ and $E_p$ are respectively the heat and the energy of reservoirs $p$, in full agreement with the first law of thermodynamics.
\end{thebibliography}
\end{document}